\documentclass[%
reprint,
groupedaddress,
nofootinbib,
 amsmath,amssymb,
 aps,
pra,
floatfix,
longbibliography
]{revtex4-1}

\usepackage{graphicx}
\usepackage{dcolumn}
\usepackage{bm}
\usepackage{braket}
\usepackage{physics}
\usepackage{comment}
\usepackage{color}
\usepackage[version=3]{mhchem}
\usepackage{ulem}

\usepackage[whole]{bxcjkjatype} 
\usepackage[colorlinks=true,urlcolor=blue,citecolor=blue,linkcolor=blue,breaklinks=true]{hyperref}

\begin{document}

\preprint{APS/123-QED}

\title{
Interaction-induced nonlinear magnon transport in noncentrosymmetric magnets}

\author{Kosuke Fujiwara}
\affiliation{Department of Applied Physics, The University of Tokyo, Hongo, Tokyo, 113-8656, Japan}

\author{Takahiro~Morimoto}
\affiliation{Department of Applied Physics, The University of Tokyo, Hongo, Tokyo, 113-8656, Japan}

\date{\today}

\begin{abstract}
We study the effect of the magnon-magnon interaction on the nonlinear magnon transport. The magnon-magnon interaction induces nonreciprocal magnon decay when the time-reversal symmetry is broken, and leads to nonlinear thermal responses of magnons. We construct a theoretical framework to study the nonlinear thermal responses due to the nonreciprocal magnon decay by using the imaginary Dyson equation and quantum kinetic theory, which is then applied to models of 1D antiferromagnets and 2D honeycomb ferromagnets with Dzyaloshinskii–Moriya interactions. An order estimate shows that the nonlinear thermal response from the present mechanism is feasible for experimental measurement.

\end{abstract}

\maketitle


\section{Introduction}
Magnons, quasiparticles of quantized spin waves, have attracted significant attention within the field of spintronics due to their long lifetimes and capability to transmit information without Joule heating~\cite{Chumak2015MagnonSpintronics}. For the practical application of magnons, it is crucial to understand the transport properties of magnons. Since the magnon is a charge-neutral boson, its thermal transport has been extensively investigated~\cite{Katsura2010TheoryMagnets,Onose2010ObservationEffect,Xiao2010TheoryEffect}, including the thermal Hall effect and the spin Nernst effect which originate from a nontrivial geometry of the magnon band~\cite{Matsumoto2011TheoreticalFerromagnets,Cheng2016SpinAntiferromagnets,Zyuzin2016MagnonAntiferromagnets}.

Beyond the linear response regime, magnons also exhibit interesting nonlinear transport phenomena. For example, driving magnons with thermal gradient leads to a nonlinear spin current, known as the nonlinear spin Seebeck effect \cite{Takashima2018NonreciprocalAntiferromagnets} and the nonlinear spin Nernst effect \cite{Kondo2022NonlinearCurrent}, where the latter emerges from the Berry curvature dipole of the magnon band. 
The application of high-intensity magnetic fields also creates magnons and leads to the DC spin current generation \cite{Proskurin2018ExcitationInsulators,Ishizuka2019TheoryInsulators,Ishizuka2022LargeTrihalides}. Furthermore, magnons in multiferroics generally have dipole moments and allow their excitation by electric fields of laser light, which was shown to be a geometric phenomenon related to the Berry connections in the case of collinear antiferromagnets \cite{Fujiwara2023NonlinearAntiferromagnets}.
The quantum kinetic theory provides a useful tool to analyze the nonlinear thermal transport of magnons \cite{Sekine2020QuantumField,Varshney2023IntrinsicApproach,Mukherjee2023NonlinearAntiferromagnets}. In this framework, the second-order nonlinear responses are divided into three parts according to their dependence on the relaxation time $\tau$ (i.e. $\propto \tau^0$, $\tau$, and $\tau^2$)~\cite{Varshney2023IntrinsicApproach}. Among them, the nonlinear Drude term, which is proportional to $\tau^2$, is particularly important when relaxation times are long. The nonzero nonlinear Drude term requires that the magnon Hamiltonian breaks the time-reversal symmetry (TRS) in addition to the inversion symmetry. 

While magnons are often treated as independent particles, magnons can interact with each other, which sometimes gives drastic changes to their properties. For example, the magnon-magnon interaction modulates the band dispersion and the lifetime of magnons~\cite{Elliott1969TheAntiferromagnets,Harris1971DynamicsHydrodynamics,Zhitomirsky1999InstabilityFields,CostaFilho2000MicroscopicProcesses,Chernyshev2006MagnonAntiferromagnets,Chernyshev2009SpinSingularities,Mourigal2010Field-inducedAntiferromagnets,Mourigal2013DynamicalAntiferromagnet,Zhitomirsky2013ColloquiumDecays,Chernyshev2016DampedFerromagnets,Maksimov2016Field-inducedAntiferromagnets,Maksimov2016Field-inducedLattice,Winter2017BreakdownMagnet,Pershoguba2018DiracFerromagnets,McClarty2018TopologicalFields,Rau2019Magnon/math,Maksimov2020Rethinking-RuCl3,Mook2020QuantumDecay,Smit2020MagnonLattice,Mook2021Interaction-StabilizedFerromagnets,Lu2021TopologicalFerromagnets,Maksimov2022Easy-planeThem,Sun2023InteractingMagnons,Koyama2023Flavor-waveModel,Chen2023DampedAntiferromagnets,Heinsdorf2023StableInteractions,Chen2023DampedAntiferromagnets,Habel2024BreakdownInsulators,Sourounis2024ImpactAntiferromagnets,Koyama2024ThermalSystems}. 
In particular, such effects on the topological magnons \cite{Chernyshev2016DampedFerromagnets,McClarty2018TopologicalFields,Mook2021Interaction-StabilizedFerromagnets,Chen2023DampedAntiferromagnets,Heinsdorf2023StableInteractions} and the associated edge modes \cite{Koyama2023Flavor-waveModel,Habel2024BreakdownInsulators}, strongly affect the magnitude of the thermal Hall effect~\cite{Mook2021Interaction-StabilizedFerromagnets,Chen2023DampedAntiferromagnets,Sourounis2024ImpactAntiferromagnets,Koyama2024ThermalSystems}. Furthermore, it was revealed that magnon-magnon interactions induce a qualitative change of quantum phases, exemplified by the interaction-induced topological magnons, where the magnon Hamiltonian breaks the time-reversal symmetry through the magnon-magnon interactions~\cite{Mook2021Interaction-StabilizedFerromagnets}.
\begin{figure}[tb]
\centering
\includegraphics[width=\linewidth]{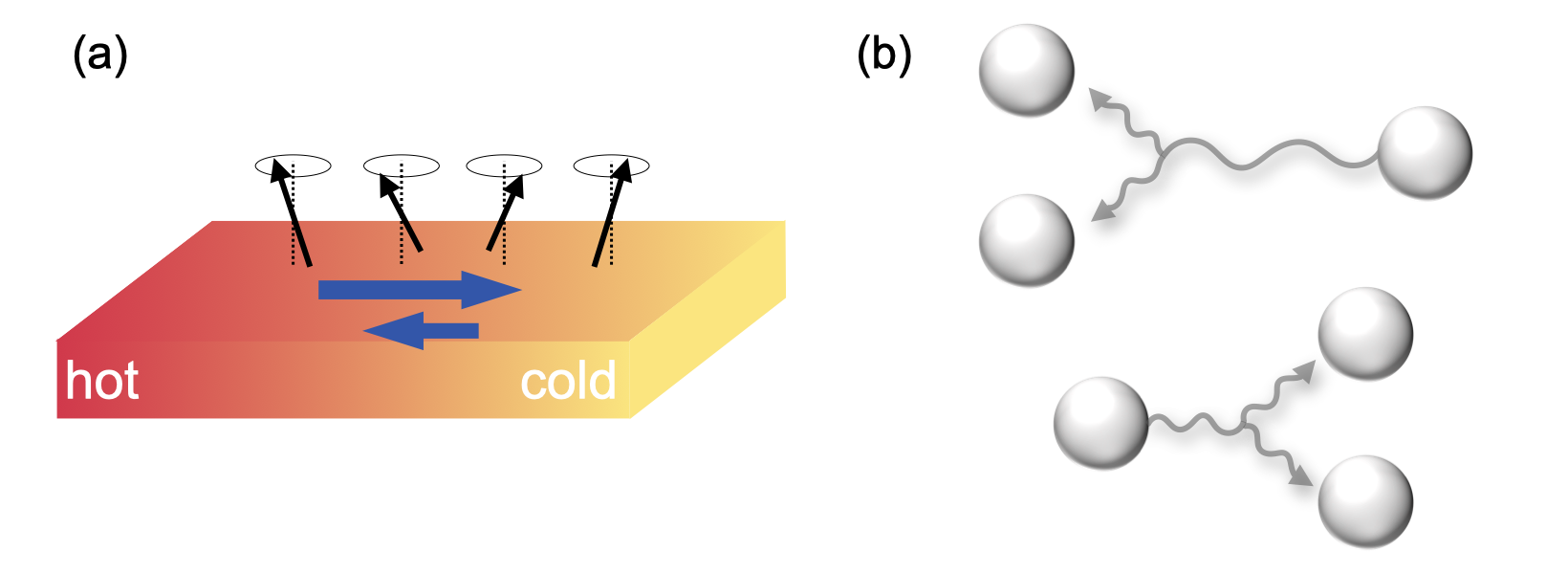}
\caption{The schematic picture of the nonreciprocal magnon thermal current and the nonreciprocal magnon decay. (a) Nonreciprocal thermal current (blue arrows) carried by magnon excitations (black arrows). (b) Nonreciprocal magnon decay arising from magnon-magnon interactions. Left-going and right-going magnons (white spheres) experience different decay rates due to inversion symmetry breaking.}
\label{fig:sketch}
\end{figure}
In this paper, we study the influence of magnon-magnon interactions on the nonlinear thermal transport of magnons and show that the magnon-magnon interaction can induce a unique nonlinear response that cannot be captured in the independent particle picture. Specifically, the magnon-magnon interaction that breaks the TRS induces nonreciprocity in the magnon decay rate, which leads to nonlinear responses of magnons (Fig.~\ref{fig:sketch}). Interestingly, even when the bilinear magnon Hamiltonian effectively preserves a TRS, the magnon-magnon interaction (Fig. ~\ref{fig:sketch}(b)) introduces breaking of the TRS and gives rise to a nonlinear Drude term. We adopt a perturbation theory and the imaginary Dyson equation to study the nonreciprocal magnon decay from the magnon-magnon interaction. This allows us to obtain the magnon lifetime and compute the nonlinear Drude terms by using the quantum kinetic theory. We apply our formalism to a model of one-dimensional antiferromagnets and a model of honeycomb ferromagnets.
We then estimate the effect of the nonlinear Drude term from the magnon-magnon interaction, which turns out to be comparable to that originating from an explicit TRS breaking for a free magnon Hamiltonian and is feasible for experimental measurements.

\section{Interaction-induced nonreciprocal magnon decay}\label{sec:method}

In this section, we present our formalism to study the effect of magnon-magnon interactions on the magnon damping rate.

Throughout this paper, we consider the spin Hamiltonian consisting of the two spin interactions along with the Zeeman interaction, which can be written as
\begin{equation}
    H_{spin}=\sum_{i,j}\frac{1}{2}\vb*{S}_{i}J_{ij}\vb*{S}_{j}-\sum_{i}\vb*{h}\cdot\vb*{S}_i.\label{spin_ham}
\end{equation}
Here, $J_{ij}$ is a matrix representing the interaction between the spin of site $i$ and the spin of site $j$.
We assume that the ground state spin configuration has a magnetic order and allows magnon expansion with respect to the realized magnetic order. 

To obtain the magnon Hamiltonian, we perform the Holstein-Primakoff transformation for the spin $S$ systems as~\cite{Holstein1940FieldFerromagnet}
\begin{subequations} 
\begin{align} \label{HPtrans} 
    S^{+}_i&=\hbar\sqrt{2S-a^\dagger_ia_i}a_i\notag\\
    &=\hbar\sqrt{2S}a_i-\hbar a^\dagger_ia_ia_i/2\sqrt{2S}+O(1/S\sqrt{S}),\\
    S^{-}_i&=\hbar a^\dagger_i\sqrt{2S-a^\dagger_ia_i}\notag\\
    &=\hbar\sqrt{2S}a^\dagger_i-\hbar a^\dagger_ia^\dagger_ia_i/2\sqrt{2S}+O(1/S\sqrt{S}),\\
    S^{z}_i&=\hbar(S-a^\dagger_i a_i),
\end{align}
\end{subequations} 
where $a^\dagger_i$ is a bosonic magnon creation operator at $i$th site, $\vb*{S}_i$ is a spin operator at $i$th site along the spin configuration of the ground state and $S^{\pm}_i=S_i^{x}\pm iS_i^{y}$. 

Up to the forth order of magnon operators, the magnon Hamiltonian
can be written as
\begin{equation}
    \mathcal{H}=E_0+\mathcal{H}_2+\mathcal{H}_3+\mathcal{H}_4,
\end{equation}
where $E_0$ is the ground state energy, $H_2$ is a bilinear Hamiltonian, $H_3$ is a cubic Hamiltonian which does not conserves magnon number, and $H_4$ is a quartic Hamiltonian. 
We ignore magnon-magnon interactions of a higher order than $H_4$ since these terms do not appear in the perturbation up to the order of $1/S$.

By using the HP transformation and the Fourier transformation on the spin Hamiltonian $H_{spin}$, we obtain the magnon Hamiltonian $\mathcal{H}_2$ as 
\begin{equation}
    \mathcal{H}_2=\frac{1}{2}\sum_{\vb*{k}} \phi^\dagger_{\vb*{k}}H_{2,\vb*{k}}\phi_{\vb*{k}}.
\end{equation}
where $\phi^\dagger_{\vb*{k}}=(a^\dagger_{\vb*{k},1}\cdots,a^\dagger_{\vb*{k},N_u},a_{-\vb*{k},1}\cdots,a_{-\vb*{k},N_u})$, $a^\dagger_{\vb*{k},\alpha}$ is the magnon creation operator of $\alpha$th band with momentum $\vb*{k}$, and $N_u$ is the number of sites in a unit cell.

Bilinear Hamiltonian $\mathcal{H}_2$ is the bosonic Bogoliubov de Gennes (BdG) Hamiltonian. Thus, we diagonalize $H_{2,\vb*{k}}$ as $H_{2,\vb*{k}}=T_{\vb*{k}}^\dagger \varepsilon_{\vb*{k}}T_{\vb*{k}}$, where $T_{\vb*{k}}$ is a paraunitary matrix which satisfies $T_{\vb*{k}}^\dagger \sigma_3T_{\vb*{k}}=\sigma_3$ and $\sigma_3=\sigma_z\otimes I_{N_u}$. 
Then $\mathcal{H}_2$ can be written as $\mathcal{H}_2=\sum_{\vb*{k}}\psi^\dagger_{\vb*{k}}\varepsilon_{\vb*{k}}\psi_{\vb*{k}}$, 
using the diagonal matrix $\varepsilon_{\vb*{k}}=\textrm{diag}(\varepsilon_{\vb*{k},1},\cdots,\varepsilon_{\vb*{k},N_u},\varepsilon_{-\vb*{k},1},\cdots,\varepsilon_{-\vb*{k},N_u})$ and
$\psi^\dagger_{\vb*{k}}=T^{-1}_{\vb*{k}}\phi(\vb*{k})=(b^\dagger_{\vb*{k},1}\cdots,b^\dagger_{\vb*{k},N_u},b_{-\vb*{k},1}\cdots,b_{-\vb*{k},N_u})$.

From the spin Hamiltonian $H_{spin}$, we also obtain interaction terms $\mathcal{H}_3$ and $\mathcal{H}_4$ as
\begin{subequations}
\begin{align}
    \mathcal{H}_3=&\frac{1}{2\sqrt{N}}\sum_{\vb*{k},\vb*{q},\vb*{p}}^{p=k+q}\sum_{\alpha,\beta,\gamma} V_{\vb*{k},\vb*{q},\vb*{p}}^{\alpha\beta\gamma}a^\dagger_{\vb*{k},\alpha}a^\dagger_{\vb*{q},\beta}a_{\vb*{p},\gamma}+h.c.\label{eq:H_3}\\
    \mathcal{H}_4=&\frac{1}{4N}\sum_{\vb*{k},\vb*{q},\vb*{p},\vb*{l}}^{\vb*{l}=\vb*{k}+\vb*{q}+\vb*{p}}\sum_{\alpha,\beta,\gamma,\delta} W_{\vb*{k},\vb*{q},\vb*{p},\vb*{l}}^{\alpha\beta\gamma\delta}a^\dagger_{\vb*{k},\alpha}a^\dagger_{\vb*{q},\beta}a^\dagger_{\vb*{p},\gamma}a_{\vb*{l},\delta}+h.c.\notag\\
    +&\frac{1}{4N}\sum_{\vb*{k},\vb*{q},\vb*{p},\vb*{l}}^{\vb*{l}=\vb*{k}+\vb*{q}-\vb*{p}}\sum_{\alpha,\beta,\gamma,\delta} Y_{\vb*{k},\vb*{q},\vb*{p},\vb*{l}}^{\alpha\beta\gamma\delta}a^\dagger_{\vb*{k},\alpha}a^\dagger_{\vb*{q},\beta}a_{\vb*{p},\gamma}a_{\vb*{l},\delta}+h.c., \label{eq:H_4}
\end{align}
\end{subequations} 
where $N$ is the number of unit cells, and $V$, $W$, and $Y$ are coefficients of the magnon-magnon interactions.

We treat the magnon-magnon interaction from $\mathcal{H}_3$ and $\mathcal{H}_4$ as a perturbation to $\mathcal{H}_2$ which we incorporate as a self-energy of a magnon \cite{Chernyshev2009SpinSingularities,Mook2021Interaction-StabilizedFerromagnets,Koyama2023Flavor-waveModel}.
Up to the order of $1/S$, the Green's function $\mathcal{G}$ in the imaginary time formalism is given by 
\begin{align}
    &\mathcal{G}_{k,\alpha\beta}(\tau)=\notag\\
    &\mathcal{G}_{k,\alpha\beta}^0(\tau)+\int^\beta_0d\tau_1\langle T_\tau \mathcal{H}_4(\tau_1)a_{k,\alpha}(\tau)a^\dagger_{k,\beta}\rangle\notag\\
    &-\frac{1}{2}\int^\beta_0d\tau_1\int^\beta_0d\tau_2\langle T_\tau \mathcal{H}_3(\tau_1)\mathcal{H}_3(\tau_2)a_{k,\alpha}(\tau)a^\dagger_{k,\beta}\rangle,\label{eq:green's_func}
\end{align}
where $\mathcal{G}^0$ is the unperturbed Green function, $\beta=1/k_BT$ is the inverse temperature, $T_\tau$ represents the imaginary time ordering of operators, and $\langle\cdots\rangle$ is the thermal average for the unperturbed Hamiltonian $\mathcal{H}_2$.
\begin{figure}[tb]
   \centering
   \includegraphics[width=\linewidth]{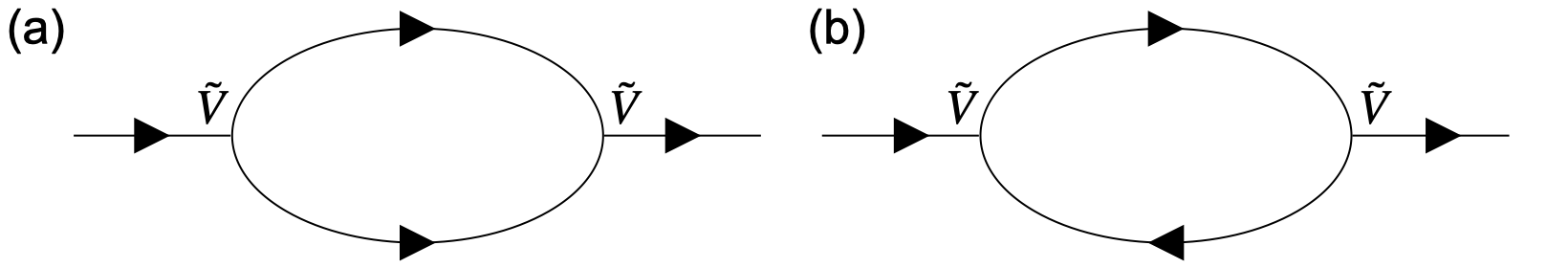}
   \caption{Diagrams of magnon-magnon interactions which contribute to the magnon damping rate in the order of $1/S$.  Bubble diagrams corresponding to (a) the first and (b) the second terms of Eq.(\ref{Sigma_k}). 
   The contribution from the diagram (a) is dominant at low temperatures.
   }
   \label{fig:diagram}
\end{figure}

The contribution of $\mathcal{H}_4$ only gives the Hartree term whose contribution is real. Consequently, it modifies the magnon energy only, leaving the magnon lifetime unchanged. Thus, we focus on the contribution of $\mathcal{H}_3$ in the following, to study the effect of magnon lifetime. We rewrite $\mathcal{H}_3$ in basis $\psi_{\vb*{k}}$, which diagonalizes the bilinear Hamiltonian $\mathcal{H}_{2,\vb*{k}}$, and we obtain
\begin{subequations}
    \begin{align}
    \mathcal{H}_3=&\frac{1}{2\sqrt{N}}\sum_{\vb*{k},\vb*{q},\vb*{p}}^{p=k+q}\sum_{\alpha,\beta,\gamma} \tilde{V}_{\vb*{k},\vb*{q},\vb*{p}}^{\alpha\beta\gamma}b^\dagger_{\vb*{k},\alpha}b^\dagger_{\vb*{q},\beta}b_{\vb*{p},\gamma}+h.c.\\
    &+\frac{1}{3!\sqrt{N}}\sum_{\vb*{k},\vb*{q},\vb*{p}}^{p=k+q}\sum_{\alpha,\beta,\gamma} \tilde{W}_{\vb*{k},\vb*{q},\vb*{p}}^{\alpha\beta\gamma}b^\dagger_{\vb*{k},\alpha}b^\dagger_{\vb*{q},\beta}b^\dagger_{\vb*{p},\gamma}+h.c.
\end{align}
\end{subequations}
The contribution of $\mathcal{H}_3$ has two types of contribution: tadpole diagrams and bubble diagrams. 
The contributions of the tadpole diagrams and the bubble diagrams of $\tilde{W}$ are real. In contrast, the bubble diagrams of $\tilde{V}$ shown in Fig.~\ref{fig:diagram} have imaginary parts which lead to the magnon decay. Thus, we focus on the contribution of the bubble diagrams to the self-energy $\Sigma$ which can be written as
\begin{align}
\Sigma_{\vb*k}^{\alpha,\beta}(\omega,T)=\frac{1}{N}\sum_q\sum_{\gamma,\gamma^\prime}\bigg(&\frac{1}{2}\frac{\tilde{V}^{\gamma,\gamma^\prime,\beta}_{\vb*{q},\vb*{k}-\vb*{q},\vb*{k}}(\tilde{V}^{\gamma,\gamma^\prime,\alpha}_{\vb*{q},\vb*{k}-\vb*{q},\vb*{k}})^*}{\omega-\varepsilon_{\vb*{q},\gamma}-\varepsilon_{\vb*{k}-\vb*{q},\gamma^\prime}+i\eta}\notag\\
&\times[f^B_{\vb*{q},\gamma,T}+f^B_{\vb*{k}-\vb*{q},\gamma^\prime,T}+1] \notag\\
&+\frac{(\tilde{V}^{\beta,\gamma,\gamma^\prime}_{\vb*{k},\vb*{q},\vb*{k}+\vb*{q}})^*\tilde{V}^{\alpha\gamma,\gamma^\prime}_{\vb*{k},\vb*{q},\vb*{k}+\vb*{q}}}{\omega+\varepsilon_{\vb*{q},\gamma}-\varepsilon_{\vb*{k}+\vb*{q},\gamma^\prime}+i\eta} \notag\\
& \times[f^B_{\vb*{q},\gamma,T}-f^B_{\vb*{k}+\vb*{q},\gamma^\prime,T}]\bigg)\label{Sigma_k}.
\end{align}
Here, $\varepsilon_{\vb*{k},\gamma}$ is a magnon energy of the band $\gamma$ determined by $\mathcal{H}_2$ and $f^B_{\vb*{k},\gamma,T}=1/(\exp(\beta\varepsilon_{\vb*{k},\gamma})-1)$ is a Bose distribution function. 
The first and second terms correspond to contributions from Fig.~\ref{fig:diagram}(a) and Fig.~\ref{fig:diagram}(b), respectively.
In particular, at the zero temperature, only the first term from Fig.~\ref{fig:diagram}(a) is nonzero and the second term from Fig.~\ref{fig:diagram}(b) vanishes, indicating that the first term (particle-particle diagram) gives a dominant contribution at low temperatures.

The non-reciprocity of the self-energy is encoded in the difference of the self-energy at the opposite momenta $k$ and $-k$.
By comparing $\Sigma_{\vb*{k}}^{\alpha,\beta}(\omega,T)$ and $\Sigma_{-\vb*{k}}^{\alpha,\beta}(\omega,T)$ (which is obtained by substituting $\vb*{k}$ to $-\vb*{k}$ and $\vb*{q}$ to $-\vb*{q}$ in Eq.~\eqref{Sigma_k}) and assuming $\varepsilon_{\vb*{k}}=\varepsilon_{-\vb*{k}}$ due to the TRS for the bilinear magnon Hamiltonian $H_{2,\vb*k}$, we find that the nonreciprocity in the self-energy $\Sigma_{\vb*{k}}^{\alpha,\beta}(\omega,T)\neq\Sigma_{-\vb*{k}}^{\alpha,\beta}(\omega,T)$ requires the condition $V^{\alpha,\gamma,\gamma^\prime}_{\vb*{k},\vb*{q},\vb*{p}}(V^{\beta,\gamma,\gamma^\prime}_{\vb*{k},\vb*{q},\vb*{p}})^*\neq V^{\alpha,\gamma,\gamma^\prime}_{-\vb*{k},-\vb*{q},-\vb*{p}}(V^{\beta,\gamma,\gamma^\prime}_{-\vb*{k},-\vb*{q},-\vb*{p}})^*$.
This condition is generally met for magnon-magnon interactions with broken TRS.

Now we focus on the magnon damping caused within each magnon band and ignore the off-diagonal part of $\Sigma$. 
We introduce the damping rate $\Gamma$ induced by the magnon-magnon interaction as the imaginary part of the self-energy 
\begin{equation}
    \Gamma_{\vb*k}^{\alpha}(\omega,T)=-\textrm{Im}\Sigma_{\vb*k}^{\alpha,\alpha}(\omega,T).
    \label{eq: Gamma k}
\end{equation}
To compute the nonlinear thermal conductivity, we use the Born approximation, which replaces $\omega$ in the right hand of Eq.~\eqref{eq: Gamma k} with $\varepsilon_{\vb*{k},\alpha}$, but this sometimes causes an unphysical divergence.
To avoid such divergence, we adopt the imaginary Dyson equation~\cite{Chernyshev2009SpinSingularities,Chernyshev2016DampedFerromagnets,Koyama2023Flavor-waveModel} at a finite temperature, in which we solve the self-consistent equation
\begin{equation}
    \tilde{\omega}=\varepsilon_{\vb*{k}}-i\Gamma_{\vb*{k}}(\tilde{\omega}^*,T),\label{iDE}
\end{equation}
where $\tilde{\omega}^*$ denotes a complex conjugate of $\tilde{\omega}$, which originates from the causality. We approximate $\Gamma_{\vb*{k}}(\omega,T)$ in Green's functions by $\Gamma_{\vb*{k}}(\tilde{\omega},T)$ and obtain the magnon damping $\Gamma_{\vb*{k}}(\tilde{\omega},T)$ by solving Eq.~\eqref{eq: Gamma k} and Eq.~\eqref{iDE} self-consistently. The magnon damping $\Gamma_{\vb*{k}}(\tilde{\omega},T)$ shows an enhancement when the magnon dispersion $\varepsilon_{\vb*{k}}$ overlaps with the two-magnon continuum $\varepsilon_{\vb*{q}}+\varepsilon_{\vb*{k}-\vb*{q}}$ or the collision continuum $\varepsilon_{\vb*{q}}-\varepsilon_{\vb*{k}+\vb*{q}}$, since the numerators in Eq.~\eqref{Sigma_k} are real for $\alpha=\beta$ and the denominators become purely imaginary. In particular, the two-magnon continuum becomes important at low temperatures because the first term of Eq.~(\ref{Sigma_k}) is nonzero even at zero temperature. In contrast, the collision becomes unimportant at low temperatures because the second term in Eq.~(\ref{Sigma_k}) vanishes at zero temperature.
If we write the magnon damping due to effects other than the magnon-magnon interaction (e.g. impurity scattering and interactions with phonons) by $\eta$, the magnon lifetime of the $\gamma$th band $\tau_{\vb*{k},\gamma}$ can be expressed as $\tau_{\vb*{k},\gamma}=1/2(\eta_{\vb*{k},\gamma}+\Gamma_{\vb*{k}}^\gamma(\tilde{\omega},T))$. Hereafter, we assume that $\eta_{\vb*{k},\gamma}=\alpha\varepsilon_{\vb*{k},\gamma}$ where $\alpha$ is a damping factor. This assumption corresponds to the phenomenological Gilbert damping.

Here we comment on the effect of the real part of the self-energy $\Sigma$. The real part of the self-energy causes asymmetric modulation of the magnon energy and velocity between $k$ and $-k$. 
Such asymmetric band modulation can affect the magnitude of the nonlinear responses. 
Generally, the calculation of the real part of the self-energy requires more computational cost than the calculation of the imaginary part alone. 
This point is discussed in Appendix~\ref{sec:appenrix_real_part}.

\section{Nonlinear thermal current of magnons}
Now we consider the nonlinear thermal current generated by the thermal gradient
\begin{equation}
    J^\mu_Q=\kappa^{\mu\nu\nu}\qty(\partial_\nu T)^2.
\end{equation}
From the quantum kinetic theory, the nonlinear Drude term is written as \cite{Varshney2023IntrinsicApproach}
\begin{align}
\kappa^{\mu\nu\nu}_{nd}=\sum_n\int \frac{dk^3}{(2\pi)^3}
&\left[-\frac{1}{\hbar T}\tau_{\vb*{k},\gamma}^2\varepsilon_{\vb*{k},\gamma}^2v^\mu_{\vb*{k},\gamma}\frac{\partial v^\nu_{\vb*{k},\gamma}}{\partial k_\nu}\frac{\partial f^B_{\vb*{k},\gamma}}{\partial T}\notag\right.\\
&\left.+\tau_{\vb*{k},\gamma}^2\varepsilon_{\vb*{k},\gamma}v^\mu_{\vb*{k},\gamma}(v^\nu_{\vb*{k},\gamma})^2\frac{\partial^2 f^B_{\vb*{k},\gamma}}{\partial^2 T}\right],
\label{eq: sigma}
\end{align}
where $v_{\vb*{k},\gamma}^\mu=\partial_{k_\mu}\varepsilon_{\vb*{k},\gamma}/\hbar$ is the group velocity of magnons in the $\gamma$th band.
If there is the TRS, $\varepsilon_{\vb*{k},\gamma}=\varepsilon_{-\vb*{k},\gamma}$, $\vb*{v}_{\vb*{k},\gamma}=-\vb*{v}_{-\vb*{k},\gamma}$, $\tau_{\vb*{k},\gamma}=\tau_{-\vb*{k},\gamma}$, the nonlinear Drude term vanishes. Thus, the nonzero nonlinear Drude term requires not only broken inversion symmetry but also broken TRS.

\section{Application }
To demonstrate nonlinear thermal current due to the nonreciprocal magnon decay, we apply our formalism to 1D antiferromagnets and 2D honeycomb ferromagnets with inversion symmetry breaking.

\subsection{1D antiferromagnets}

\begin{figure}[tb]
\centering
\includegraphics[width=\linewidth]{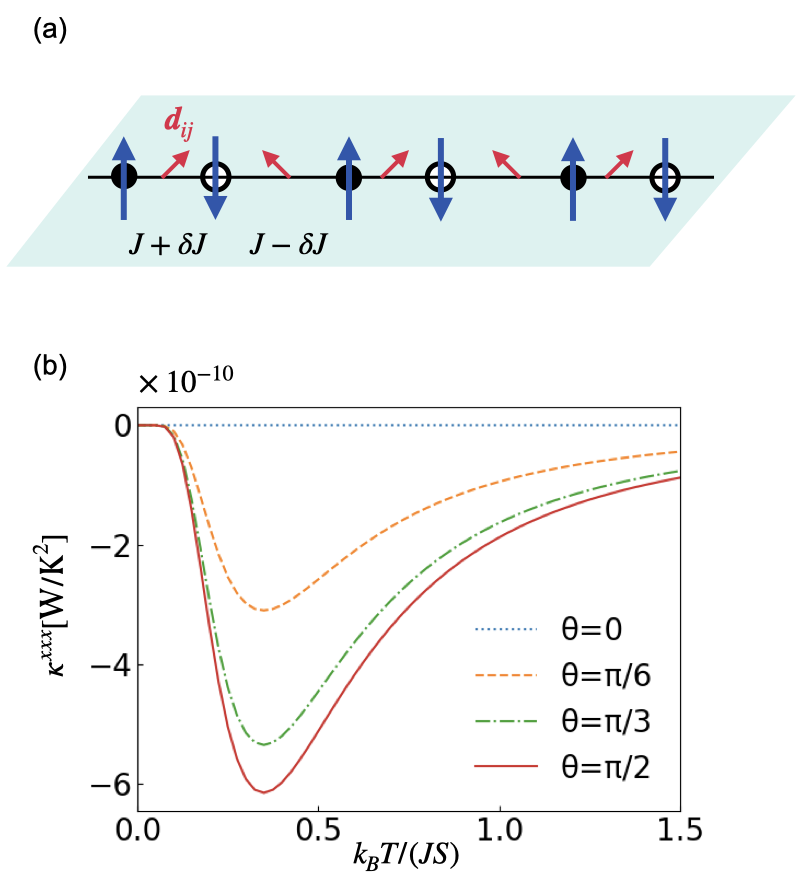}
\caption{Nonlinear thermal transport in the 1D inversion broken antiferromagnet. (a) The spin model of the 1D antiferromagnet with DM interaction. Blue arrows represent spin moments and red arrows represent DM interaction between the neighboring spins.
(b) DM interaction dependence of $\kappa^{xxx}$. 
We used the following parameters: $\delta J/J=0.2$, $S=1$, $h/JS=0.2$, $\alpha=0.001$, $\vb*{d}_1/J=(0.1,0.1,0.0)/\sqrt{2}$ and $\vb*{d}_2$ is a $\theta$ rotation of $\vb*{d}_1$ around the $z$ axis.}
\label{fig:1d_model}
\end{figure}

We study thermal current in one-dimensional antiferromagnets. We consider the  model with broken inversion symmetry induced by the alternating exchange interaction and Dzyaloshinskii–Moriya interaction (DM interaction) in the direction perpendicular to the antiferromagnetic spin order, as depicted in Fig. \ref{fig:1d_model}. The Hamiltonian is given by
\begin{align}
    H=\sum_{i}&\bigg[(J+(-1)^i\delta J)(S_i^xS_{i+1}^x+S_i^yS_{i+1}^y+\alpha S_i^zS_{i+1}^z)\notag\\
    &+\vb*{d}_{1}\cdot(\vb*{S}_{2i}\times\vb*{S}_{2i+1})+\vb*{d}_{2}\cdot(\vb*{S}_{2i+1}\times\vb*{S}_{2i+2})\notag\\
    &-hgS^z_i \bigg],\label{eq.1d_anti}
\end{align}
where $\vb*{S}_i$ is a spin at $i$th site and we assume that the ground state spin configuration is given by $\vb*{S}_i =(-1)^i\vb*{z}$ ($\vb*{z}$ is the unit vector along the z direction). $\vb*{d}_{m}$ ($m=1,2$) are DM vector oriented in the $xy$ plane. We note that the DM interaction gives zero contribution to the ground state energy since $\vb*{S}_i\times\vb*{S}_j=0$ in the ground state configuration, but the DM interaction affects the excitations.

The bilinear magnon Hamiltonian for \eqref{eq.1d_anti} is written as
\begin{equation}
    H_{2,k}=2S
    \begin{pmatrix}
    \alpha J+\frac{gh}{2S} & J\cos{ka}-i\delta J\sin{ka}\\
    J\cos{ka}+i\delta J\sin{ka} & \alpha J-\frac{gh}{2S}
    \end{pmatrix},
\end{equation}
where $a$ is a lattice constant. The cubic Hamiltonian $H_3$ is induced by the DM interaction and non-zero components of $V^{abc}_{\vb*{k},\vb*{q},\vb*{p}}$ in Eq.(\ref{eq:H_3}) are
\begin{subequations} 
\begin{align}
    V^{1,2,1}_{\vb*{k},\vb*{q},\vb*{p}}&=\sqrt{\frac{S}{2}}(-d_{1y}-id_{1x})\exp(-iqa),\\
    V^{1,2,2}_{\vb*{k},\vb*{q},\vb*{p}}&=\sqrt{\frac{S}{2}}(-d_{1y}+id_{1x})\exp(ika),\\
    V^{2,1,2}_{\vb*{k},\vb*{q},\vb*{p}}&=\sqrt{\frac{S}{2}}(d_{2y}-id_{2x})\exp(-iqa),\\
    V^{2,1,1}_{\vb*{k},\vb*{q},\vb*{p}}&=\sqrt{\frac{S}{2}}(d_{2y}+id_{2x})\exp(ika).
\end{align}
\end{subequations} 
If the orientation of $\vb*{d}_{1}$ and $\vb*{d}_{2}$ is parallel or anti-parallel, the state is invariant by the time-reversal operation and $\pi$ rotation of spins about $\vb*{d}_{m}$. Thus, the system has the effective TRS and the magnon Hamiltonian has TRS. However, when the orientation of $\vb*{d}_{m}$ depends on bonds, the DM interaction breaks the effective TRS, resulting in magnon-magnon interactions that break TRS of the magnon Hamiltonian. By calculating $\tau_{\vb*{k}}$ and using Eq.~\eqref{eq: sigma}, we obtain the nonlinear Drude term as shown in Fig. \ref{fig:1d_model} (b). 

The blue dotted curve in Fig. \ref{fig:1d_model} (b) shows the result for the bond-independent DM interaction $(\vb*{d}_{1} = \vb*{d}_{2})$.
In this case, the nonlinear thermal conductivity vanishes as $\kappa^{xxx}=0$ because of the effective TRS, even when a DM interaction and a 3-magnon interaction are present. 
When the DM interaction has a bond-dependence with a finite $\theta$ ($\theta$: the angle between $\vb*{d}_1$ and $\vb*{d}_2$), the 3-magnon interaction breaks the TRS of the magnon Hamiltonian, resulting in a nonreciprocal response as shown by the orange, green, and red curves for different values of $\theta$ in Fig. \ref{fig:1d_model} (b). 
It can be seen that the response increases as $\theta$ increases in the range of $0$ to $\pi/2$. 
We note that incorporating the effect of the real part of $\Sigma$ modifies the result for $\kappa^{xxx}$ quantitatively, but does not lead to a qualitative change (For details, see Appendix \ref{sec:appenrix_real_result}).

\subsection{Honeycomb ferromagnets \label{sec: honeycomb}}

\begin{figure*}[tb]
\centering
\includegraphics[width=\linewidth]{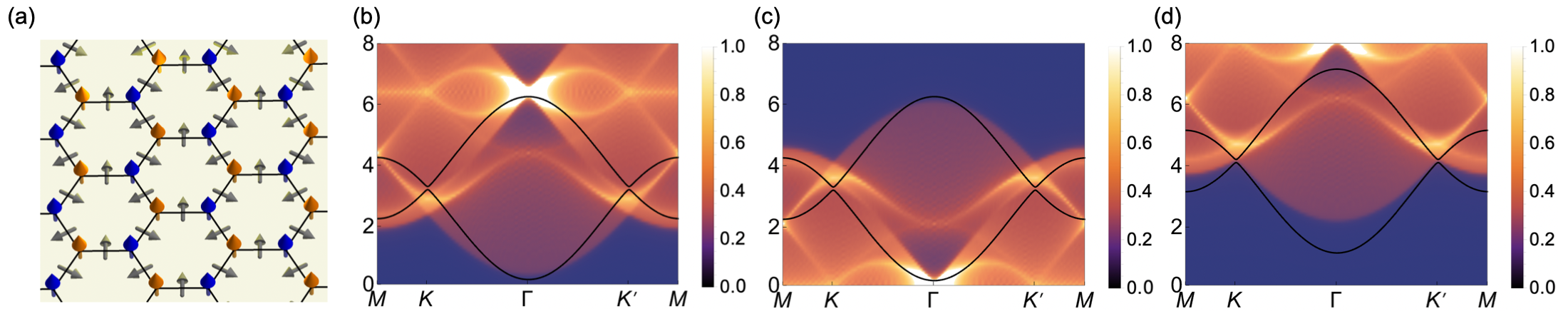}
\caption{The schematic picture of the spin model and the associated band dispersion. (a) The honeycomb lattice ferromagnets with DM interaction. Orange and blue arrows represent spin moments and gray arrows represent DM interaction between the neighboring spins. (b-d) The magnon band dispersion (black curves) along high-symmetry paths of the Brillouin zone. The color plot shows $D_{two}$ ((b) and (d)) and $D_{coll}$ (c) with the maximum value being normalized to $1$. $\Delta_A/J=0.0$, $\Delta_B/J=0.05$, $\mu_A=\mu_B=1.0$, $S=1$ and $h/JS=0.1$ for (b) and (c) and $h/JS=1.0$ for (d).}
\label{fig:model}
\end{figure*}

Next, we consider the spin model of two-dimensional ferromagnets on the honeycomb lattice as depicted in Fig. \ref{fig:model} (a). The spin Hamiltonian is given by
\begin{align}
    H=&-\frac{J}{2}\sum_{\langle i,j \rangle}\vb*{S}_i\cdot\vb*{S}_{j}+\frac{D}{2}\sum_{\langle i,j \rangle}\vb*{d}_{ij}\cdot\vb*{S}_i\times\vb*{S}_{j} \nonumber \\
    &-h\sum_i g_{\alpha} S^z_i-\sum_i \Delta_\alpha(S^z_i)^2, \label{Hamiltonian}
\end{align}
where $\vb*{d}_{ij}=\vb*{z}\times(\vb*{r}_j-\vb*{r}_i)/|\vb*{r}_j-\vb*{r}_i|$ is a DM vector oriented in the $xy$ plane, and $g_{\alpha}$ is the $g$ factor for the spins on $\alpha~(\alpha=A,B)$ sites and $\Delta_\alpha$ is the magnetic anisotropy for $\alpha$ sites. Here, we consider the symmetry of the spin Hamiltonian. In ferromagnets, the spin configuration breaks the TRS because the time-reversal flips spins. However, there still exists an effective TRS, composed of time-reversal and rotation in spin space, in the absence of the DM interaction ($D=0$)\cite{Mook2021Interaction-StabilizedFerromagnets}. This model also possesses an effective inversion-symmetry, consisting of inversion and rotation in spin space, in the absence of the magnetic anisotropy difference ($\Delta_A=\Delta_B$).
Therefore, a nonzero nonlinear Drude term requires $D \neq 0$ and $\Delta_A \neq \Delta_B$ by breaking both effective TRS and inversion symmetry.
Moreover, this honeycomb magnet model has a $C_3$ rotation symmetry and the symmetry $IM_x$ composed of inversion symmetry and mirror symmetry along the $yz$-plane. Due to the $C_3$ rotation symmetry the nonlinear Hall current due to the Berry curvature dipole ($\propto \tau$) vanishes \cite{Sodemann2015QuantumMaterials}, and the nonlinear thermal responses satisfy $\kappa^{yyy}=-\kappa^{yxx}$ and $\kappa^{xxx}=-\kappa^{xyy}$. Also, the symmetry $IM_x$ leads to $\kappa^{xxx}=0$. Therefore, the nonlinear thermal response in this model arises from the nonlinear Drude term and satisfies $\kappa^{yyy}=-\kappa^{yxx}$ and $\kappa^{xxx}=\kappa^{xyy}=0$.

\begin{figure}[tb]
\centering
\includegraphics[width=\linewidth]{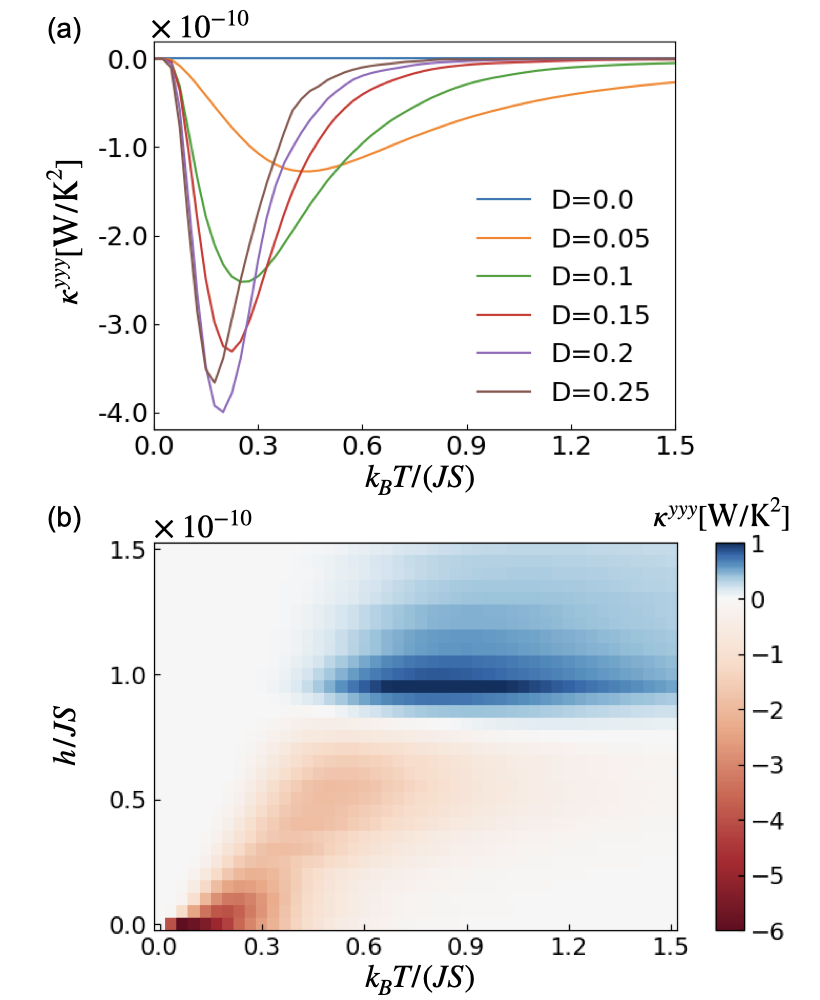}
\caption{Nonlinear thermal conductivity $\kappa^{yyy}$ induced by the nonreciprocal magnon decay. (a) Temperature dependence of $\kappa^{yyy}$ for different values of the DM interaction. (b) The color plot of $\kappa^{yyy}$. $\kappa^{yyy}$ shows a sign change around $h/JS=1.0$. We used the following parameters: $\Delta_A/J=0.0$, $\Delta_B/J=0.05$, $\mu_A=\mu_B=1.0$, $S=1$ and $\alpha=0.001$. We used $h/JS=0.1$ for (a), $D/J=0.15$ for (b).} 
\label{fig:results}
\end{figure}

The bilinear magnon Hamiltonian $ H_{2,\vb*{k}}$ for Eq.~\eqref{Hamiltonian} is written as
\begin{equation}
    H_{2,\vb*{k}}=
    \begin{pmatrix}
    3JS+hg_A+2S\Delta_A & -JS\gamma_{\vb*{k}}\\
    -JS\gamma_{-\vb*{k}} & 3JS+hg_B+2S\Delta_B
    \end{pmatrix},
\end{equation}
with $\gamma_{\vb*{k}}=\sum_{j=1}^3e^{i\vb*{k}\cdot\vb*{\delta}_j}$, where $\delta_j$ are the vectors pointing from one site to three neighboring sites. We show the magnon band dispersion and the two-magnon density of states $D_{two}(\omega)=\frac{1}{N}\sum_{\gamma,\gamma^\prime}\sum_{\vb*{q}\in{BZ}}\delta(\omega-\varepsilon_{\vb*{q},\gamma}-\varepsilon_{\vb*{k}-\vb*{q},\gamma^\prime})$ in Fig.~\ref{fig:model}(b) and (d) and the magnon band dispersion and the collision density of states $D_{coll}(\omega)=\frac{1}{N}\sum_{\gamma,\gamma^\prime}\sum_{\vb*{q}\in{BZ}}\delta(\omega-\varepsilon_{\vb*{q},\gamma}+\varepsilon_{\vb*{k}+\vb*{q},\gamma^\prime})$ in Fig.~\ref{fig:model}(c). As we mentioned before, 
when the magnon dispersion $\varepsilon_{\vb*{k}}$ overlaps with the region of large $D_{two}$ and $D_{coll}$, the magnon-magnon interaction has a significant effect on the magnon lifetime.
Now, we incorporate the effect of the $H_3$ that breaks the effective TRS as a magnon damping by solving Eq.~\eqref{iDE}. An explicit expression for $H_3$ is given in Appendix \ref{sec:appenrix_interaction}.
By using Eq.~\eqref{eq: sigma}, we obtain the nonlinear Drude term as shown in Fig. \ref{fig:results}.
We note that incorporating the effect of the real part of $\Sigma$ does not change the result much as shown in Appendix~\ref{sec:appenrix_real_result}. 

Figure \ref{fig:results} (a) shows the nonlinear thermal conductivity $\kappa^{yyy}$ for six values of the DM interaction. When there is no DM interaction ($D=0$), the nonlinear Drude term vanishes ($\kappa^{yyy}=0$). For the finite DM interactions, nonvanishing $\kappa^{yyy}$ appears and the peak of $\kappa^{yyy}$ shifts to lower temperatures as the DM interaction increases. 
This behavior can be explained by the two-magnon continuum and the collision continuum. From Fig.~\ref{fig:model}(b), we can see that at low energies, the overlap between $\varepsilon_{\vb*{k}}$ and the two-magnon continuum is small. Thus, when the DM interaction is small and $H_3$ is small, the contribution of the first term in Eq.~(\ref{Sigma_k}) to $\Gamma_{\vb*{k}}$ is small. While the magnon dispersion $\varepsilon_{\vb*{k}}$ shows a significant overlap with the collision continuum at low energy, as shown in Fig.~\ref{fig:model}(c), the contribution of the second term in Eq.~(\ref{Sigma_k}) to $\Gamma_{\vb*{k}}$ is small at low temperatures as we mentioned before. Thus, when the DM interaction is small, $\kappa^{yyy}$ has a peak at high temperatures. 
On the other hand, as the DM interaction increases, the contribution of the first term in Eq.~(\ref{Sigma_k}) to $\Gamma_{\vb*{k}}$ increases even if the overlap between $\varepsilon_{\vb*{k}}$ and the two-magnon continuum is not so large. This results in the shift of the peak of $\kappa^{yyy}$ toward lower temperatures, as the DM interaction increases.
Furthermore, up to $D/J\sim0.2$, the magnitude of $\kappa^{yyy}$ increases as the DM interaction increases, but above $D/J\sim0.2$, the magnitude of $\kappa^{yyy}$ decreases as the DM interaction increases (Fig.~\ref{fig:results}(a)). Specifically, the nonreciprocity in the magnon decay increases for larger DM interaction,
whereas,
for too large DM interaction, the lifetime of magnons becomes short and the nonlinear response proportional to $\tau^2$ is suppressed. It has been noted that as the interaction becomes stronger, the magnon band is expelled from the continuum and the magnon lifetime increases \cite{Verresen2019AvoidedInteractions}. However, in this parameter region, such phenomena only occur in regions where $D/J$ is much larger than $0.25$ (see Appendix~\ref{sec:appenrix_lifeteime_interaction}).

Figure \ref{fig:results} (b) shows the magnetic field and the temperature dependence of the nonlinear thermal conductivity $\kappa^{yyy}$. As the magnetic field increases, the response appears at the higher temperature, and furthermore, the sign of the response changes around $h/JS\sim1$. This behavior can be explained by the energy shift and overlap between the magnon energy $\varepsilon_{\vb*{k}}$ and the two-magnon continuum or the collision continuum. Due to the magnetic field, the magnon band dispersion shifts to higher energy. In particular, if $\mu_A=\mu_B=\mu$, magnon energy shifts by $\mu h$, while the two-magnon continuum shifts by $2\mu h$. 
In particular, in the range $h/JS > 1$, only the upper magnon band overlaps with the two-magnon continuum as shown in Fig.~\ref{fig:model}(d). 
Since the sign of magnon group velocity $\vb*{v}_{\vb*{k}}$ is opposite between the lower and upper bands, the sign of the nonlinear thermal conductivity changes around $h/JS \simeq 1$.
Since the collision continuum is independent of the magnetic field, an overlap of magnon energy with the collision continuum becomes very small for large magnetic fields, resulting in less contribution to the nonlinear conductivity.
In general, as the magnetic field is increased, the overlap between the magnon energy and the two-magnon continuum and collision continuum becomes smaller and the effect of magnon-magnon interaction becomes smaller.

\section{Discussion}
Let us estimate the order of magnitude of the nonlinear thermal current induced by the magnon-magnon interaction. Assuming that the lattice constant is $4$~\AA~and the interlayer distance is $6$~\AA~ by taking the values for $\ce{CrX_3}$ ($\ce{X}=\ce{Br},~\ce{I}$)~\cite{McGuire2017CrystalTrihalides}, we obtain the nonlinear thermal conductivity of $\sigma \simeq 5 \times 10^{-10}$ 
$\textrm{W}/\textrm{K}^2$ from Fig.~\ref{fig:results}(b).
For the temperature gradient $\nabla T \simeq 10^5$ K/m, the nonlinear thermal current of $J_Q \simeq 5$ W/m$^2$.
In particular, the nonlinear Hall thermal current from $\kappa^{yxx}$ is feasible for experimental measurements.
Since the typical value of linear thermal Hall conductivity is given by $\kappa^{xy} \simeq 10^{-4}\sim10^{-3}$ W/Km~\cite{Onose2010ObservationEffect,Ideue2012EffectInsulators,Hirschberger2015ThermalMagnet}, the linear contribution gives $J_Q \simeq 10\sim100$ W/m$^2$ for the temperature gradient $\nabla T \simeq 10^5$ K/m. Therefore, the nonlinear contribution to the thermal Hall current from $\kappa^{yxx}$ is sizable compared to the linear contribution from $\kappa^{xy}$.

In this study, we adopted the spin model which gives the harmonic magnon Hamiltonian $H_2$ with the effective TRS, while the magnon-magnon interaction breaks the effective TRS and induces the nonreciprocal magnon damping and the nonlinear responses. In general models, $H_2$ can also break the effective TRS (which we refer to here as the ``explicitly broken TRS'') and induce a nonlinear response within the harmonic theory. 
Let us compare the magnitudes of the nonlinear responses induced by the magnon-magnon interaction and the ``explicitly broken TRS''. Here, we consider a model where the DM vector is oriented in the $z$ direction and $H_2$ does not have the effective TRS in a similar way to the Haldane model. In this ``explicitly broken TRS'' model, we obtain $\kappa^{\mu\nu\nu}$ of the order of $10^{-10}$~W/K$^2$ which is the same order as the nonlinear thermal conductivity induced by the magnon-magnon interaction (for details see Appendix \ref{sec:appenrix_comparison}), indicating that the effect of the magnon-magnon interaction is not negligible even in cases with the explicitly broken TRS. 
In addition, the non-reciprocity from these two mechanisms shows qualitatively different behaviors with respect to the DM interaction $D$.
The lifetime of the magnon incorporating the magnon-magnon interaction is written as
\begin{align}
    \tau_{\vb*{k},D}&=1/(\alpha\varepsilon_{\vb*{k}}+\Gamma_{\vb*{k}})\sim1/(\alpha\varepsilon_{\vb*{k}}+c_1\varepsilon_{\vb*{k}}D^2/J^2)\notag\\
    &\sim \tau_{\vb*{k},D=0}(1-c_1 D^2/J^2\alpha),
\end{align}
whereas that for explicitly broken TRS systems without the magnon-magnon interaction is
\begin{align}
    \tau_{\vb*{k},D}&=1/(\alpha\varepsilon_{\vb*{k},D})\sim1/\alpha\varepsilon_{\vb*{k},D=0}(1+c_2 D/J)\notag\\
    &\sim \tau_{\vb*{k},D=0}(1-c_2D/J).
\end{align}
Here, $c_1$ and $c_2$ are coefficients of order of unity.
Therefore, if $D/J$ is larger than $\alpha$, the non-reciprocity induced by the magnon-magnon interaction becomes important compared to that induced by the ``explicitly broken TRS''.
In addition, the in-plane DM interaction is expected to be larger than the perpendicular interaction in the honeycomb model, given that the in-plane DM interaction arises from the nearest neighbor interactions, whereas the perpendicular DM interaction for the ``explicitly broken TRS'' arises from the next-nearest neighbor interaction. Therefore, the in-plane DM interaction can be larger than the perpendicular DM interaction, where the magnon-magnon interaction can contribute more significantly to the nonlinear thermal response.

The multiferroic kamiokite materials $M\ce{_2Mo_3O_8}$ ($M$:3d transition metal)~\cite{Ideue2012EffectInsulators,Park2020ThermalInteraction} can be the candidate material for the magnon-magnon interaction-induced nonlinear responses. One may also consider a heterostructure of honeycomb ferromagnets $\ce{CrI_3}$ or $\ce{CrBr_3}$~\cite{Yelon1971RenormalizationEffects,Chen2018TopologicalCrI3,Burch2018MagnetismMaterials} on top of a substrate to introduce inversion symmetry breaking to the system. While we considered the in-plane DM interaction as the origin of the magnon-magnon interaction, the Kitaev-$\Gamma$ model~\cite{Xu2018InterplayMonolayers,Lee2020FundamentalCrI3,Nakazawa2022AsymmetricField} also gives rise to the magnon-magnon interactions and application of our theory leads to the nonreciprocal magnon decay in a similar way.
Antiferromagnetic materials are also candidate materials. While we studied honeycomb ferromagnets in this paper, honeycomb antiferromagnets can also give rise to non-reciprocal bands due to in-plane DM interactions~\cite{Sourounis2024ImpactAntiferromagnets}. Candidate materials for such antiferromagnets on honeycomb lattices include $\ce{MnPS_3}$, $\ce{MnPSe_3}$, and $\ce{VPS_3}$~\cite{Bazazzadeh2021SymmetryMonolayers,Xing2019MagnonAntiferromagnets}. In particular, non-reciprocal magnon bands have been investigated in $\ce{MnPS_3}$~\cite{Wildes2021SearchMnPS3}.

Our theory is also applicable to noncollinear magnetic materials. In the noncollinear case, quantum corrections due to 3-magnon interactions can change the ground state~\cite{Zhitomirsky1999InstabilityFields}, and this correction should be taken into account.

\begin{acknowledgments}
We thank fruitful discussions with Alexander Mook, Sota Kitamura, Shun Okumura, Joji Nasu, Shinnosuke Koyama, Hosho Kastura.
T.M. was supported by 
JSPS KAKENHI Grant 23K25816, 23K17665, 24H02231, and
JST CREST (Grant No. JPMJCR19T3).
K.F. was supported by JSPS KAKENHI Grant 24KJ0730 and the Forefront Physics and Mathematics program to drive transformation (FoPM).
\end{acknowledgments}

\appendix

\section{Real part of self-energy}\label{sec:appenrix_real_part}
\begin{figure}[tb]
   \centering
   \includegraphics[width=\linewidth]{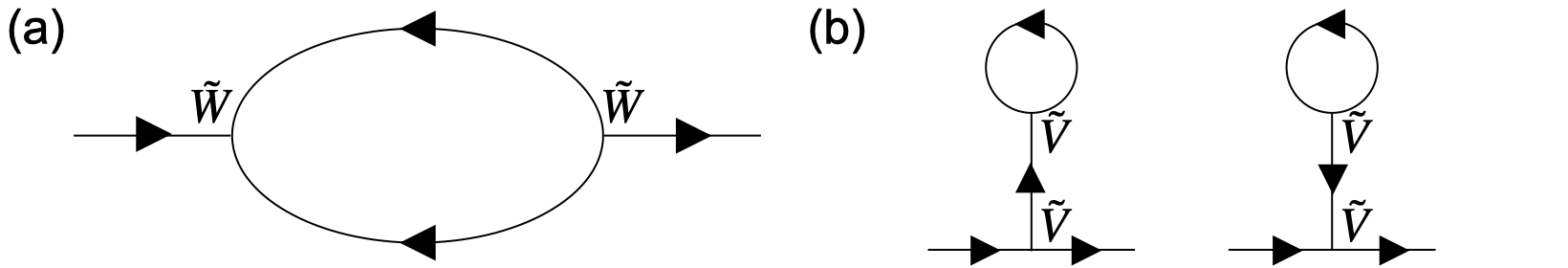}
   \caption{Diagrams of magnon-magnon interactions which contribute to the magnon energy.  Bubble diagrams corresponding to (a) $\Sigma_W$ depicted in Eq.~\eqref{Appendixeq:Sigma_W}, and
   (b) $\Sigma_{tad}$ depicted in Eq.~\eqref{Appendixeq:Sigma_tad}.}
   \label{appendixfig:diagram}
\end{figure}

In Sec. \ref{sec:method}, we consider the imaginary part of the self-energy depicted as Eq.~(\ref{Sigma_k}). However, self-energy written as Eq.~\eqref{eq:green's_func} has other terms that contribute to the real part of the self-energy $\mathrm{Re}\Sigma$: bubble diagram containing $\tilde{W}^{\alpha\beta\gamma}_{\vb*{k},\vb*{q},\vb*{p}}$ depicted in Fig.~\ref{appendixfig:diagram}(a), tadpole diagram depicted in Fig.~\ref{appendixfig:diagram}(b) and the first order perturbation of $H_4$. 
In this section, we consider the effect of $\mathrm{Re}\Sigma$ originating from these terms. 

The contribution of the bubble diagram of $\tilde{W}^{\alpha\beta\gamma}_{\vb*{k},\vb*{q},\vb*{p}}$ and the tadpole diagram is
\begin{align}
    \Sigma^{\alpha,\beta}_{W,\vb*{k}}(\omega,T)=\frac{1}{2N}\sum_q\sum_{\gamma,\gamma^\prime}&\frac{\tilde{W}^{\gamma,\gamma^\prime,\alpha}_{\vb*{q},-\vb*{k}-\vb*{q},\vb*{k}}(\tilde{W}^{\gamma,\gamma^\prime,\beta}_{\vb*{q},-\vb*{k}-\vb*{q},\vb*{k}})^*}{\omega+\varepsilon_{\vb*{q},\gamma}+\varepsilon_{-\vb*{k}-\vb*{q},\gamma^\prime}-i\eta}\notag\\
    &\times\big(f^B_{\varepsilon_{\vb*{q},\gamma},T}+f^B_{\varepsilon_{-\vb*{k}-\vb*{q},\gamma^\prime},T}+1\big),\label{Appendixeq:Sigma_W}
\end{align}
\begin{align}
    \Sigma^{\alpha,\beta}_{tad,\vb*{k}}(\omega,T)=\frac{1}{N}\sum_{\vb*{q}}\sum_{\gamma,\delta}(&(\tilde{V}^{\beta\gamma\alpha}_{\vb*{k},\vb*{0},\vb*{k}})^*\tilde{V}^{\gamma\delta\delta}_{\vb*{0},\vb*{q},\vb*{q}}\notag\\
    &+(\tilde{V}^{\gamma\delta\delta}_{\vb*{0},\vb*{q},\vb*{q}})^*\tilde{V}^{\alpha\gamma\beta}_{\vb*{k},\vb*{0},\vb*{k}})\frac{f^B_{\varepsilon_{\vb*{q},\delta},T}}{\varepsilon_{\vb*{0},\gamma}}.\label{Appendixeq:Sigma_tad}
\end{align}
The contribution of $\Sigma^{\alpha,\beta}_{W,\vb*{k}}$ and $\Sigma^{\alpha,\beta}_{tad,\vb*{k}}$ cause nonreciprocal band renormalization since $H_3$ breaks TRS. Therefore, the real part of $\Sigma^{\alpha,\beta}_{W,\vb*{k}}$ and $\Sigma^{\alpha,\beta}_{tad,\vb*{k}}$ can contribute to the nonlinear Drude term.
To incorporate the real part of the self-energy, we consider the total self-energy $\Sigma^{\alpha,\beta}_{tot,\vb*{k}}(\omega,T)=\Sigma^{\alpha,\beta}_{\vb*{k}}(\omega,T)+\Sigma^{\alpha,\beta}_{W,\vb*{k}}(\omega,T)+\Sigma^{\alpha,\beta}_{tad,\vb*{k}}(\omega,T)$ and off-shell Dyson equation~\cite{Chernyshev2009SpinSingularities,Chernyshev2016DampedFerromagnets,Koyama2023Flavor-waveModel}
\begin{equation}
    \tilde{\omega}=\varepsilon_{\vb*{k}}+\Sigma_{tot,\vb*{k}}(\tilde{\omega}^*,T),\label{Appendixeq:dyson}.
\end{equation} 
Then, we obtain the renormalized magnon energy $\textrm{Re}[\tilde{\omega}]$ and the magnon damping $\textrm{Im}[\tilde{\omega}]$. Here, the calculation of the real part requires principal-valued integration, which leads to a convergence behavior worse than that with the imaginary part only.

The contribution of $H_4$ is incorporated as an effective Hamiltonian
\begin{align}
H_{eff}=&\frac{1}{4N}\sum_{\vb*{k},\vb*{q},\vb*{p},\vb*{l}}^{\vb*{l}=\vb*{k}+\vb*{q}+\vb*{p}}\sum_{\alpha,\beta,\gamma,\delta} W_{\vb*{k},\vb*{q},\vb*{p},\vb*{l}}^{\alpha\beta\gamma\delta}\bigg(\ev{a^\dagger_{\vb*{k},\alpha}a^\dagger_{\vb*{q},\beta}}a^\dagger_{\vb*{p},\gamma}a_{\vb*{l},\delta}\notag\\
&+\ev{a^\dagger_{\vb*{p},\gamma}a_{\vb*{l},\delta}}a^\dagger_{\vb*{k},\alpha}a^\dagger_{\vb*{q},\beta}+\ev{a^\dagger_{\vb*{k},\alpha}a^\dagger_{\vb*{p},\gamma}}a^\dagger_{\vb*{q},\beta}a_{\vb*{l},\delta}\notag\\
&+\ev{a^\dagger_{\vb*{q},\beta}a_{\vb*{l},\delta}}a^\dagger_{\vb*{k},\alpha}a^\dagger_{\vb*{p},\gamma}+\ev{a^\dagger_{\vb*{k},\alpha}a_{\vb*{l},\delta}}a^\dagger_{\vb*{q},\beta}a^\dagger_{\vb*{p},\gamma}\notag\\
&+\ev{a^\dagger_{\vb*{q},\beta}a^\dagger_{\vb*{p},\gamma}}a^\dagger_{\vb*{k},\alpha}a_{\vb*{l},\delta}
\bigg)+h.c.\notag\\
+&\frac{1}{4N}\sum_{\vb*{k},\vb*{q},\vb*{p},\vb*{l}}^{\vb*{l}=\vb*{k}+\vb*{q}-\vb*{p}}\sum_{\alpha,\beta,\gamma,\delta} Y_{\vb*{k},\vb*{q},\vb*{p},\vb*{l}}^{\alpha\beta\gamma\delta}\bigg(\ev{a^\dagger_{\vb*{k},\alpha}a^\dagger_{\vb*{q},\beta}}a_{\vb*{p},\gamma}a_{\vb*{l},\delta}\notag\\
&+\ev{a_{\vb*{p},\gamma}a_{\vb*{l},\delta}}a^\dagger_{\vb*{k},\alpha}a^\dagger_{\vb*{q},\beta}+\ev{a^\dagger_{\vb*{k},\alpha}a_{\vb*{p},\gamma}}a^\dagger_{\vb*{q},\beta}a_{\vb*{l},\delta}\notag\\
&+\ev{a^\dagger_{\vb*{q},\beta}a_{\vb*{l},\delta}}a^\dagger_{\vb*{k},\alpha}a_{\vb*{p},\gamma}+\ev{a^\dagger_{\vb*{k},\alpha}a_{\vb*{l},\delta}}a^\dagger_{\vb*{q},\beta}a_{\vb*{p},\gamma}\notag\\
&+\ev{a^\dagger_{\vb*{q},\beta}a_{\vb*{p},\gamma}}a^\dagger_{\vb*{k},\alpha}a_{\vb*{l},\delta}\bigg)+h.c.
\label{appendixeq: Heff}
\end{align}
In our model, the contribution of $H_{eff}$ causes a reciprocal band renormalization since spin Hamiltonian only contains the two-body interaction of spins and $H_4$ has the same symmetry as the $H_2$. $H_{eff}$ has two components, a thermal one, which is proportional to the magnon number and finite only at finite temperature, and a quantum one, which is finite at zero temperature~\cite{Sourounis2024ImpactAntiferromagnets}. Ferromagnetic models only contain the thermal term, and hence, the contribution of $H_4$ vanishes at zero temperature.
In general, writing $H_{eff}$ in terms of a basis that diagonalizes $H_2$ can result in negative energy. To avoid this unphysical situation, it is necessary to calculate $H_{eff}$ in a basis incorporating the band renormalization effect of $H_4$. Specifically, obtaining a physically valid $H_{eff}$ necessitates to find a basis diagonalizing $H_2+H_{eff}$ in a self-consistent way. 

\section{Effect of the band energy modulation $\mathrm{Re}\Sigma$}\label{sec:appenrix_real_result}

\begin{figure}[tb]
    \centering
    \includegraphics[width=\linewidth]{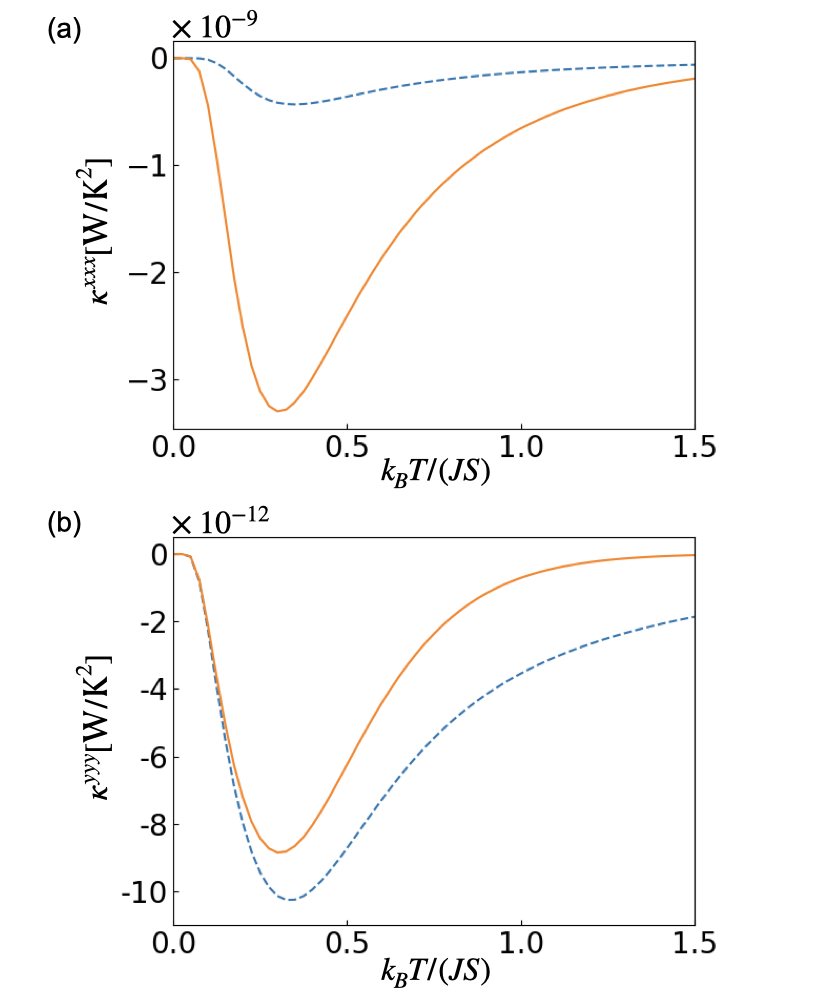}
    \caption{Nonlinear Drude terms calculated with the real part of the self-energy for (a) one-dimentional antiferromagnetic model and (b) honeycomb ferromagnetic model. The blue dashed curves represent the nonlinear Drude term when only the imaginary part of the self-energy is considered, and the orange solid curves represent the nonlinear Drude term when the real part of the self-energy is considered. We used the following parameters for (a): $\delta J/J=0.2$, $S=1$, $h=0.2$, $\alpha=0.001$, $\vb*{d}_1/J=(0.1,0.1,0.0)/\sqrt{2}$ and $\vb*{d}_2=(-0.1,0.1,0.0)/\sqrt{2}$. We used the following parameters for (b): $\Delta _A/J=0.0$, $\Delta _B/J=0.05$, $\mu_A=\mu_B=1.0$, $D/J=0.2$, $S=1$, $h/JS=0.1$ and $\alpha=0.01$.}
    \label{appendixfig:real_part}
 \end{figure}

 In this section, we calculate the nonlinear Drude term with the real part of the self-energy for the 1D antiferromagnetic model and the honeycomb ferromagnetic model. As mentioned in Appendix~\ref{sec:appenrix_real_part}, due to the bad convergence of the principal-valued integrals, a relatively large phenomenological damping is included in the two-dimensional system.  We also consider a parameter region where negative energy does not appear without self-consistently obtaining a basis diagonalizing $H_2+H_{eff}$, and instead, we use the basis  diagonalizing $H_2$ for obtaining $H_{eff}$ to reduce the numerical cost.
 
The results for the nonlinear Drude term are shown in Fig.~\ref{appendixfig:real_part}.
The nonlinear Drude terms with $\mathrm{Re}\Sigma$ are shown in orange curves and those without $\mathrm{Re}\Sigma$ are shown in blue dashed curves for comparison.
The results do not change qualitatively when the real part of the self-energy is taken into account, both in the 1D antiferromagnetic model and in the honeycomb ferromagnetic model. Here, we note that the nonlinear response of the two-dimensional model shown in Fig.~\ref{appendixfig:real_part}(b) is much smaller than the response shown in Fig.~\ref{fig:results} because we use large phenomenological damping for numerical stability. However, quantitatively, the magnitude of the response changes by a factor of about 5 for the 1D antiferromagnetic model, whereas it does not change much for the honeycomb ferromagnetic model.
This is because the 1D antiferromagnetic model has a larger real part of the self-energy, due to nonzero contributions from $\tilde{W}$ in Fig.~\ref{appendixfig:diagram}(a), the tadpole diagram in Fig.~\ref{appendixfig:diagram}(b), and $H_{eff}$ in Eq.~\eqref{appendixeq: Heff}. 
On the other hand, in the honeycomb ferromagnetic model, the contribution of $H_{eff}$ is zero at zero temperature and the contributions of $\tilde{W}$ and the tadpole diagram are zero at any temperature. Therefore, the effect of the real part of the self-energy is small in the honeycomb ferromagnetic model.

\section{Magnon-magnon interactions for ferromagnetic Heisenberg model}\label{sec:appenrix_interaction}
We present an expression for the cubic Hamiltonian $H_3$ in Eq. (\ref{eq:H_3}). In the ferromagnetic Heisenberg model on the honeycomb model defined by Eq. (\ref{Hamiltonian}), $H_3$ is introduced by the in-plane DM interaction and the non-zero components of $V^{abc}_{\vb*{k},\vb*{q},\vb*{p}}$ in Eq.(\ref{eq:H_3}) are
\begin{subequations} 
\begin{align}
    V^{1,2,1}_{\vb*{k},\vb*{q},\vb*{p}}&=-D\sqrt{\frac{S}{2}}\sum_{j=1}^{3}e^{i\phi_{\delta_j}-i\vb*{q}\cdot\vb*{\delta}_j,}\\
    V^{1,2,2}_{\vb*{k},\vb*{q},\vb*{p}}&=D\sqrt{\frac{S}{2}}\sum_{j=1}^{3}e^{i\phi_{\delta_j}+i\vb*{k}\cdot\vb*{\delta}_j},\\
    V^{2,1,1}_{\vb*{k},\vb*{q},\vb*{p}}&=-D\sqrt{\frac{S}{2}}\sum_{j=1}^{3}e^{i\phi_{\delta_j}-i\vb*{k}\cdot\vb*{\delta}_j},\\
    V^{2,1,2}_{\vb*{k},\vb*{q},\vb*{p}}&=-D\sqrt{\frac{S}{2}}\sum_{j=1}^{3}e^{i\phi_{\delta_j}+i\vb*{q}\cdot\vb*{\delta}_j},
\end{align}
\end{subequations} 
where $\phi_{\delta_j}=\arg{(\vb*{d}_{\delta_j}^y-i\vb*{d}_{\delta_j}^x)}$ and $\vb*{d}_{\delta_j}$ is the direction of the DM interaction on the bond $\vb*{\delta}_j$ written as
\begin{subequations}
\begin{align}
    \vb*{d}_{\delta_1}&=(0,1),\\
    \vb*{d}_{\delta_2}&=(-\frac{\sqrt{3}}{2},-\frac{1}{2}),\\
    \vb*{d}_{\delta_3}&=(\frac{\sqrt{3}}{2},-\frac{1}{2}).
\end{align}
\end{subequations}

\section{Magnon lifetime for strong interactions}\label{sec:appenrix_lifeteime_interaction}

\begin{figure}[htb]
\centering
\includegraphics[width=\linewidth]{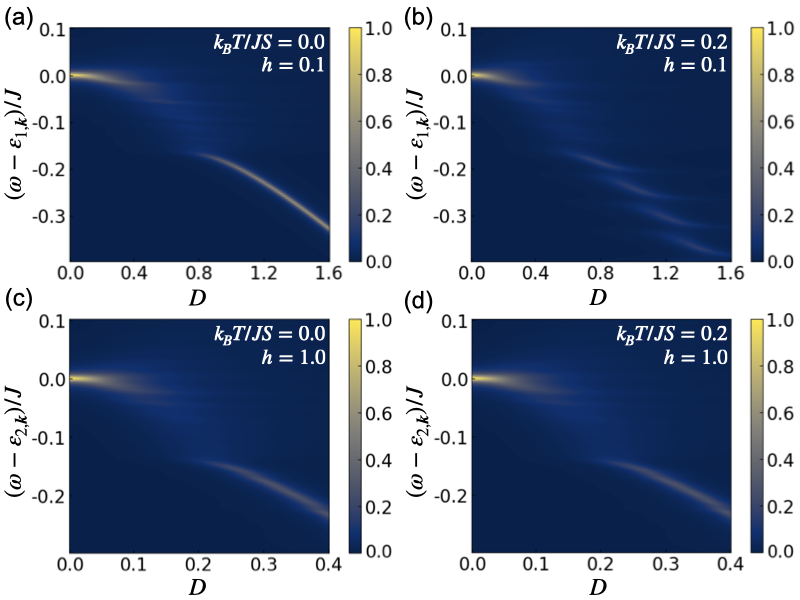}
\caption{The spectral function of the honeycomb ferromagnetic model as a function of the strength of the DM interaction $D$. In the color plot, we show the spectral function with the maximum value being normalized to $1$. 
(a,b) The spectral function for $\vb*{k}=0.4\times(2\pi/3, 2\pi/(3\sqrt{3}))$ ($(2\pi/3, 2\pi/(3\sqrt{3}))$ corresponds to the $K$ point) and $h/JS=0.1$ at (a) the zero temperature and (b) the finite temperature $k_BT/JS=0.2$. (c,d) The spectral function for $\vb*{k}=0.9\times(2\pi/3, 2\pi/(3\sqrt{3}))$ and $h/JS=1.0$ at (c) the zero temperature and (d) the finite temperature $k_BT/JS=0.2$. We used the following parameters:  $\Delta_A/J=0.0$, $\Delta_B/J=0.05$, $\mu_A=\mu_B=1.0$, $S=1$, $\alpha=0.01$.} 
\label{appendixfig:D_dep}
\end{figure}

We showed that the overlap between the magnon band and the continuum is important for the large magnon danping in Sec.~\ref{sec: honeycomb}. It is worth noting that the strong interaction repels the magnon band from the continuum and the magnon band no longer overlaps with the continuum~\cite{Verresen2019AvoidedInteractions}. In other words, under strong interactions, the magnon band does not overlap with the continuum and a long lifetime can be obtained. 
Here, we show that a long lifetime can be obtained at very strong DM interaction ($D/J \sim 0.8 \gg 0.25$) at the parameter range used in Fig~\ref{fig:results} (a). 
Meanwhile, at the finite temperature, the collision continuum makes a major contribution to the magnon damping where the magnon does not necessarily obtain a long lifetime.

Here, we consider the magnon lifetime for the honeycomb ferromagnetic model with the strong magnon-magnon interaction. Figure.~\ref{appendixfig:D_dep} shows the spectral function for several parameters.
We consider the magnon band with $h/J=0.1$ (see Fig.~\ref{fig:model}(b) and (c)). At the zero temperature and $\vb*{k}=0.4(2\pi/3, 2\pi/(3\sqrt{3}))$ ($(2\pi/3, 2\pi/(3\sqrt{3}))$ corresponds to the $K$ point), the lower magnon band is repelled from the two-magnon continuum and the spectral function become sharp as shown in Fig.~\ref{appendixfig:D_dep}(a). However, at the finite temperature, the lower magnon band has a finite lifetime as shown in Fig.~\ref{appendixfig:D_dep}(b), since the magnon band overlaps with the collision continuum. For the larger magnetic field $h/J=1.0$ (see Fig.~\ref{fig:model}(d)), the upper magnon band and the collision continuum do not overlap with each other. Therefore, the spectral function becomes sharp when the interaction strength exceeds about $D/J=0.2$ regardless of the temperature as shown in Fig.~\ref{appendixfig:D_dep} (c) and (d).

Comparing the low magnetic field case ($h/J=0.1$) with the high magnetic field case ($h/J=1.0$) at zero temperature, it is clear that the low magnetic field case requires a larger interaction for the spectral function to become sharper.
This can be understood from the density of the two-magnon continuum.
The shift of the real part of poles of spectral function $A(\omega)$ is proportional to $D_{two}(\omega)$~\cite{Verresen2019AvoidedInteractions}. Thus, if $D_{two}(\varepsilon_{\vb*{k}})$ is small, the repulsion of the magnon band and two-magnon continuum is small. Therefore, the magnon band has an overlap with the continuum and the magnon lifetime is short even for the large magnon-magnon interactions. On the other hand, if $D_{two}(\varepsilon_{\vb*{k}})$ is large, the magnon band is repelled from the continuum and the magnon lifetime becomes long. 

For the parameters used in Fig.~\ref{fig:results} (a), the repulsion of the magnon band and continuum is not so important, because the low energy part (where $D_{two}$ is not so large) has a large contribution to the nonlinear response as shown in Fig.~\ref{appendixfig:D_dep} (a). 
Also, at the finite temperature, even if the magnon band no longer overlaps with the two-magnon continuum, the magnon still has a lifetime due to the magnon-magnon interaction because of the contribution of the collision continuum as shown in Fig.~\ref{appendixfig:D_dep} (b). Thus, the nonreciprocal magnon damping is important even for the large magnon-magnon interaction.
In contrast, the situation becomes different for larger magnetic fields, where the regions with larger $D_{two}$ become important. In such cases, the magnon band and continuum will no longer overlap for large magnon-magnon interactions, and the interaction-induced nonreciprocal damping may become smaller.

\section{The ``explicitly broken TRS'' model}\label{sec:appenrix_comparison}

\begin{figure}[htb]
\centering
\includegraphics[width=\linewidth]{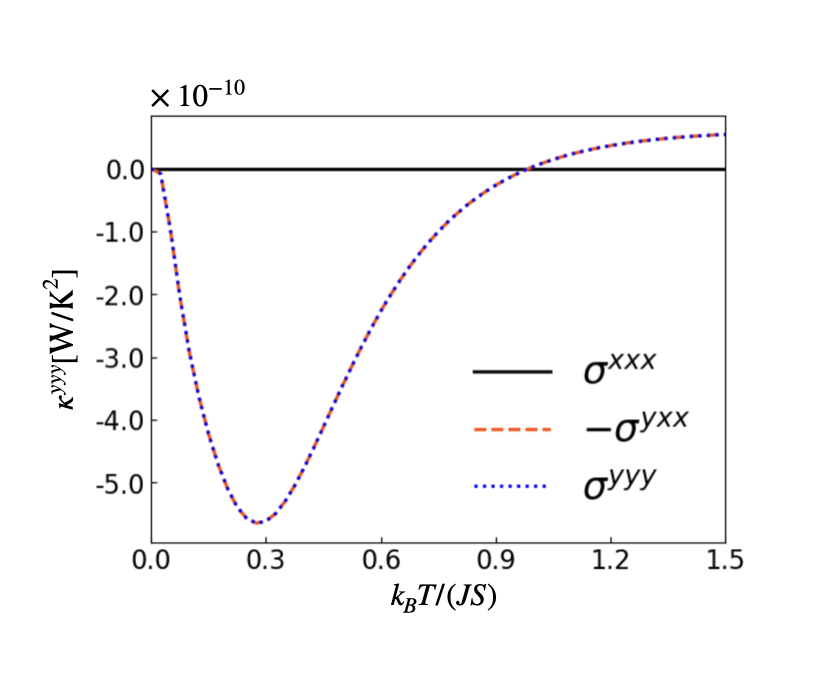}
\caption{Temperature dependence of $\kappa^{\mu\nu\nu}$ calculated with the ``explicitly broken TRS'' model $H_z$. We used the following parameters: $D/J=0.15$, $\Delta_A=0.0$, $\Delta_B=0.05$, $h/JS=0.1$, $\mu_A=\mu_B=1.0$, $S=1$ and $\alpha=0.001$.} 
\label{fig:comparison}
\end{figure}

We consider the order of the magnitude of the nonlinear thermal responses of a system with ``explicitly broken TRS''. We write the Hamiltonian of this model as $H_z$ which is the Hamiltonian (\ref{Hamiltonian}) with the modification $\vb*{d}_{ij}=(0,0,D)$ for the next-nearest neighbor bond for $A$ sites and $\vb*{d}_{ij}=(0,0,-D)$ for the next-nearest neighbor bond for $B$ sites. In this model, because of the broken effective TRS and inversion symmetry, the nonlinear Drude term can be nonzero within the harmonic theory with $H_2$. This model also has the $C_3$ rotation symmetry and the symmetry $IM_x$ composed of inversion symmetry and mirror symmetry along the $yz$-plane. Thus, we have $\kappa^{yyy}=-\kappa^{yxx}$ and $\kappa^{xxx}=\kappa^{xyy}=0$. We assume that the magnon lifetime is determined by the phenomenological damping $\alpha$ as $\tau_{\vb*{k}}=1/2\alpha\varepsilon_{\vb*{k}}$ and we obtain the nonlinear conductivity $\kappa^{\mu\nu\nu}$ for the ``explicitly broken TRS'' model as shown in Fig.~\ref{fig:comparison} by using the same parameters as Fig.~\ref{fig:results}(a). By using the peak value of the $\kappa^{\mu\nu\nu}$, we assume that the order of $\kappa^{\mu\nu\nu}$ is $10^{-10}$~W/K$^2$ which is the same order as the nonlinear thermal conductivity induced by the magnon-magnon interaction. Thus, the magnon-magnon interaction is comparable to the ``explicitly broken TRS'' of the free magnon Hamiltonian.

\nocite{*}
\bibliography{references}

\end{document}